\newcommand{\CLASSINPUTtoptextmargin}{2cm}
\newcommand{\CLASSINPUTbottomtextmargin}{3cm}
\documentclass[journal,10pt,twocolumn]{IEEEtran}
\usepackage{cite}
\usepackage{diagbox}
\usepackage{amsmath,amssymb,amsfonts}
\usepackage{algorithmic}
\usepackage{graphicx,color}
\usepackage{textcomp}
\usepackage{xcolor}
\usepackage{hyperref}
\hypersetup{hidelinks=true}
\usepackage{subfigure}
\usepackage{graphicx}
\usepackage{algorithm,algorithmic}
\def\BibTeX{{\rm B\kern-.05em{\sc i\kern-.025em b}\kern-.08emT\kern-.1667em\lower.7ex\hbox{E}\kern-.125emX}}
\AtBeginDocument{\definecolor{tmlcncolor}{cmyk}{0.93,0.59,0.15,0.02}\definecolor{NavyBlue}{RGB}{0,86,125}}

\usepackage[font=small,labelfont=bf]{caption}
\def\BibTeX{{\rm B\kern-.05em{\sc i\kern-.025em b}\kern-.08em
    T\kern-.1667em\lower.7ex\hbox{E}\kern-.125emX}}

\def\authorrefmark#1{\ensuremath{^{\textbf{#1}}}}
\newcommand\blfootnote[1]{%
  \begingroup
  \renewcommand\thefootnote{}\footnote{#1}%
  \addtocounter{footnote}{-1}%
  \endgroup
}

\begin{document}
\markboth{}{Tian, Pjanić {et al.}: Attention-aided Outdoor Localization In Commercial 5G NR Systems}

\title{Attention-aided Outdoor Localization in Commercial 5G NR Systems}

\author{Guoda Tian\authorrefmark{1,*}, Member, IEEE, Dino Pjanić\authorrefmark{1,2,*},  Student Member, IEEE, \\Xuesong Cai\authorrefmark{1}, Senior Member, IEEE,  \\Bo Bernhardsson\authorrefmark{3},  and Fredrik Tufvesson\authorrefmark{1}, Fellow, IEEE\\
$^1$ Dept. of Electrical and Information Technology, Lund University, Sweden \\
$^2$ Ericsson AB, Sweden\\
$^3$ Dept. of Automatic Control, Lund University, Sweden \\
$^*$ Equal contribution}
%\corresp{Corresponding authors: Guoda Tian (email: guoda.tian@eit.lth.se) and Xuesong Cai (email: xuesong.cai@eit.lth.se)}
\maketitle
\blfootnote{This work has been funded by Ericsson AB, the Swedish Foundation for Strategic Research, and partly by the Horizon Europe Framework Programme under the Marie Skłodowska-Curie grant agreement No. 101059091. \\}
\begin{abstract}
The integration of high-precision cellular localization and machine learning (ML) is considered a cornerstone technique in future cellular navigation systems, offering unparalleled accuracy and functionality.  This study focuses on localization based on uplink channel measurements in a fifth-generation (5G) new radio (NR) system.  %In this paper, we tackle uplink cellular massive MIMO localization tasks and the base-station operates in beam space. 
An attention-aided ML-based single-snapshot localization pipeline is presented, which consists of several cascaded blocks, namely a signal processing block, an attention-aided block, and an uncertainty estimation block. %an attention-embedded network, and an uncertainty estimation block.  
Specifically, the signal processing block generates an impulse response beam matrix for all beams. The attention-aided block trains on the channel impulse responses using an attention-aided network, which captures the correlation between impulse responses for different beams. The uncertainty estimation block predicts the probability density function of the UE position, thereby also indicating the confidence level of the localization result. Two representative uncertainty estimation techniques, the negative log-likelihood and the regression-by-classification techniques, are applied and compared. Furthermore, for dynamic measurements with multiple snapshots available, we combine the proposed pipeline with a Kalman filter to enhance localization accuracy. To evaluate our approach, we extract channel impulse responses for different beams from a commercial base station. The outdoor measurement campaign covers Line-of-Sight (LoS), Non Line-of-Sight (NLoS), and a mix of LoS and NLoS scenarios. The results show that sub-meter localization accuracy can be achieved. 
\end{abstract}


\maketitle
\section{INTRODUCTION}
\IEEEPARstart{R}{adio}-based positioning is envisioned to pave the way for numerous sophisticated yet practical applications, including vehicle navigation, intelligent traffic management, and autonomous driving \cite{THzMagazine, Russ, Indoornav, Vehiclenav, Indoorrobot, Russnav, Survey}. In contemporary 5G systems, there is a pronounced demand for precise localization capabilities. Currently, most localization-aware applications are facilitated by Global Navigation Satellite Systems (GNSS). However, the effectiveness of these systems is limited by many factors, such as shadowing, multipath propagation, and clock drifts between the GNSS transmitter and receiver \cite{GNSS2}. Consequently, there is an increasing need to investigate cellular-based technologies and seamlessly integrate those techniques into existing localization systems.

Existing cellular-based localization methods can be broadly classified into two categories, namely conventional signal processing methods \cite{Survey, 2014Double, IoTACF, Xuhong, Bayesian, 2017Direct, xuesong2, xuesong3}, and machine learning (ML) based methods \cite{GP2,TVT1,TVT2, Guoda1,Joao, Attention, Probfuse, IEEEAccess, TAP}.  
Conventional signal processing methods, such as Time of Arrival (ToA), Angle of Arrival (AoA), and Time Difference of Arrival (TDoA), require the estimation of essential channel parameters, such as signal propagation time between UE and base stations (BS). In the next step, the location of the user equipment (UE) can be estimated using these parameters. Although some of these methods have reached maturity, they can be constrained by calibration needs and algorithmic complexities \cite{Survey}. On the other hand, ML methods present a promising solution but require access to data for training and a radio environment with enough unique features that can be learned. To implement an ML-based localization approach, the initial step involves obtaining various channel fingerprints, such as the raw transfer function \cite{Probfuse, Attention}, received signal strength \cite{GP2},  angle-delay spectrum \cite{Joao, TVT1, IEEEAccess} and/or covariance matrix \cite{Guoda1, TVT2}. These fingerprints then serve as input for the ML algorithms. It should be noted that an effective method of combining several different fingerprints has the potential to significantly increase the localization accuracy, see \cite{Guoda1, TVT2}. ML-based localization algorithms can also be divided into two categories, namely classical ML approaches such as K-nearest neighbors (KNN) \cite{Guoda1}, Gaussian process regression \cite{GP2}, adaptive boosting \cite{TVT2}, and deep learning based approaches, such as fully connected neural networks (FCNN) \cite{IEEEAccess, Probfuse}, convolutional neural network \cite{GP1,GP2,TVT2,TAP}, and attention-aided networks \cite{Attention}. In particular, the attention-aided approach holds significant promise, as its embedded attention mechanism enables ML algorithms to recognize relationships between different input feature vectors, irrespective of their actual spatial or temporal separation among those vectors. This mechanism is also the core of widely used transformer techniques, producing fruitful results in various domains such as language translations, image recognition, and speech recognition \cite{Attention_google}. Another crucial aspect for localization is uncertainty prediction, which is particularly important in life-critical tasks such as autonomous driving. This research problem has been initially tackled by previous works \cite{Guoda1, Maximilian}, which provide not only the estimated location coordinates but also the corresponding variances using the negative log-likelihood (NLL) loss function.

However, to the best of our knowledge, there are still notable research gaps. Primarily, the application of attention-aided localization algorithms in 5G new radio (NR) systems represents a novel, yet unexplored, area. Secondly, the NLL uncertainty estimation technique assumes a Gaussian distribution for the estimation error of the UE position. However, such an assumption often diverges from reality. Consequently, it becomes crucial to explore further uncertainty estimation methods capable of estimating distributions other than Gaussian. To address the issues stated above, we propose a novel localization pipeline and evaluate it using data from a commercial 5G NR base station. Very few studies in the literature have been conducted on commercial grade 5G NR systems. Our research contributions are listed as follows\footnote{Initial outdoor UE localization results have been presented in the conference paper \cite{ICC}. In the current study, we utilize a higher subchannel resolution of the UL SRS channel estimates and a high-accuracy GNSS receiver. Furthermore, we apply an improved localization pipeline.}:
\begin{itemize}
    \item We apply attention-aided neural networks as the backbone to perform localization, we also demonstrate the advantages of this network in terms of localization accuracy.
    \item We apply a novel regression-by-classification method that can predict the uncertainty of localization estimates. Compared with the NLL approach, this approach provides better uncertainty estimation since it is not bounded by the assumption of Gaussianity. 
    \item We further enhance localization accuracy by applying a Kalman filter to exploit temporal correlation between multiple channel snapshots, which smoothes the estimated trajectory.
    \item Finally, we verify the novel ML-powered pipeline with real measurement data obtained using a commercial 5G NR test setup, covering both Line-of-Sight (LoS) and non-Line-of-Sight (NLoS) scenarios. The results show that our approach achieves submeter-level localization accuracy. 
\vspace{-5pt}    
    %by conducting an outdoor measurement campaign covering both line-of-sight (LoS) and non-line-of-sight (NLoS) scenarios in a commercial 5G NR test setup. It is noteworthy that our approach achieves sub-meter-level positioning accuracy.
\end{itemize}

%\IEEEPARstart{C}{ellular}-based localization is expected to pave the way for various location-aware applications such as robotic navigation, emergency healthcare, and smart transportation \cite{THzMagazine, Russ, Indoornav, Vehiclenav, Indoorrobot, Russnav, Survey}.  

The remainder of this paper is organized as follows. Section \ref{Section2} introduces the signal model and discusses the selected fingerprints. In Section \ref{Section3}, we elaborate on the localization algorithms. Section \ref{Section4} illustrates the measurement campaign and Section \ref{Section5} presents the results. Finally, conclusive remarks are included in Section \ref{Section6}. 

\section{System model and dataset generation}
\label{Section2}
We consider a commercial 5G NR system in a single-user massive MIMO scenario, where the BS processes uplink~(UL) Sounding Reference Signal (SRS) data.  The system utilizes orthogonal frequency division multiplexing (OFDM) with $F$ subcarriers, and the SRS data is a time series of uplink (UL) measurements in the beam domain. With this approach, we essentially capture the angular delay spectrum of the radio channel, an approach that has been shown to be advantageous for accurate localization based on ML \cite{Joao, Russ2}. %At time~$t$, the UE transmits an uplink pilot signal with two active antennas. %The pilot signal arrives at the BS with an arrival azimuth angle $\phi$ and an elevation angle $\theta$. 
The BS is equipped with $M_{\text{BS}}$ antenna ports, half of which is vertically polarized and the other half horizontally polarized, while the UE is equipped with $M_{\text{UE}}$ antenna ports. We suppose that the number of multipath components is $P$, and denote $\tau_{p,t}$ as the time delay between UE and BS w.r.t. the $p$-th path at time $t$, and $\alpha_{p,m,t}$ indicates the complex coefficient of each multipath component. The BS utilizes all vertical-polarized antennas to formulate $N_{\text{V}}$ beams, the response of the $i$-th beam w.r.t. the $p$-th path is $\beta_{\text{V},i}(\phi_p, \theta_p, f)$, where $f$ denotes frequency, and $\phi_p$ and $\theta_p$ represent the azimuth and elevation arrival angles for the $p$-th multipath, respectively. Similarly, another $N_{\text{H}}$ set of beams uses all horizontal polarized antennas, and the response of the $i$-th beam is $\beta_{\text{H},i}(\phi_p, \theta_p,f)$. %The total number of beams is $N = N_{V} +  N_{H}$.  
For the $m$-th UE port,  the propagation channel model for each beam at time index $t$ can be formulated as 
\begin{equation}
\begin{aligned}
     h_{\text{V},i,m,t}(f) &= \sum_{p=1}^P \beta_{\text{V},i}(\phi_p, \theta_p, f)\hspace{1pt}\alpha_{p,m,t}\hspace{1pt}\exp\{-j2\pi\hspace{1pt}f\hspace{1pt}\tau_{p,t}\} \\
      h_{\text{H},i,m,t}(f) &= \sum_{p=1}^P \beta_{\text{H},i}(\phi_p, \theta_p, f)\hspace{1pt}\alpha_{p,m,t}\hspace{1pt}\exp\{-j2\pi\hspace{1pt}f\hspace{1pt}\tau_{p,t}\}.
\end{aligned}
\end{equation}
By collecting all $h_{\text{V},i,m,t}(f)$ and $ h_{\text{H},i,m,t}(f)$ for the $F$ subcarriers, we can formulate two beam space matrices of the channel transfer function (CTF), $\mathbf{H}_{\text{V},m,t} \in \mathbb{C}^{N_{V} \times F}$ and $\mathbf{H}_{\text{H},m,t} \in \mathbb{C}^{N_{\text{H}} \times F}$ at time $t$, which correspond to the vertical and horizontal polarized antenna groups, respectively. We further define matrix $\mathbf{H}_t \in \mathbb{C}^{N \times F} = \left[\mathbf{H}_{\text{H},1,t}^T,\mathbf{H}_{\text{V},1,t}^T,..., \mathbf{H}_{\text{H},M_{\text{UE}},t}^T,\mathbf{H}_{\text{V},M_{\text{UE}},t}^T\right]^T$ that combines channel matrices of all UE antenna ports. Specifically, $N = M_{\text{UE}}\hspace{1pt}(N_{\text{H}}+N_{\text{V}})$. 
This matrix depends strongly on the UE position, therefore they can be selected as raw channel fingerprints to perform ML-based localization.
\section{The ML-based localization approach}
\begin{figure}[t!]
	\centering  %图片全局居中
	%\vspace{-1.00cm} %设置与上面正文的距离
    \includegraphics[width=\linewidth]{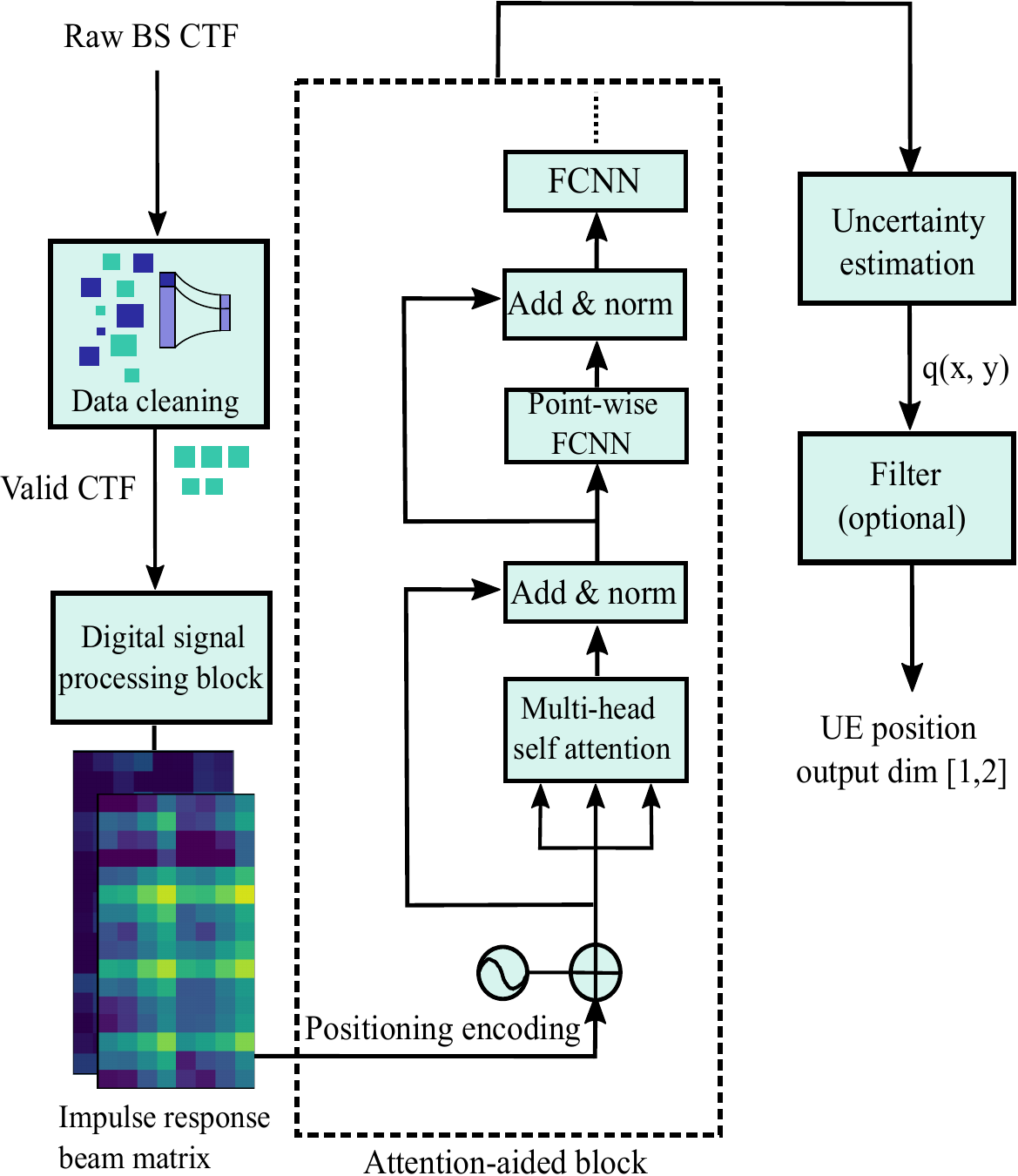}%NNCostupdate\
    \vspace{5pt}
	\caption{The ML-based localization pipeline for the 5G NR system.}
    \vspace{-10pt}%which consists an input layer, several hidden layers and an output layer.}
	\label{Pipeline} 
\end{figure}
The ML-based localization pipeline, as described in Fig.~\ref{Pipeline}, consists of five sequential blocks. First, the raw CTF $\mathbf{H}_t$ is fed into a data cleaning block to evaluate the validity of the input data. After this, valid CTFs are forwarded to a digital signal processing block to generate an impulse response beam matrix $\mathbf{G}_t \in \mathbb{C}^{N \times F}$. The amplitudes in this matrix then serve as input to a deep neural network, which incorporates a self-attention mechanism at its core. The network's final layer outputs an estimated probability density function (PDF) representing the location, thereby facilitating uncertainty estimation. To further enhance localization accuracy, a filter may be applied after the final layer of the pipeline, provided that information from multiple snapshots is available.

\label{Section3}

\subsection{The attention mechanism}%{-3pt}
\begin{figure}
 %\begin{subfigure}
  \centering
  \includegraphics[width=0.90\linewidth]{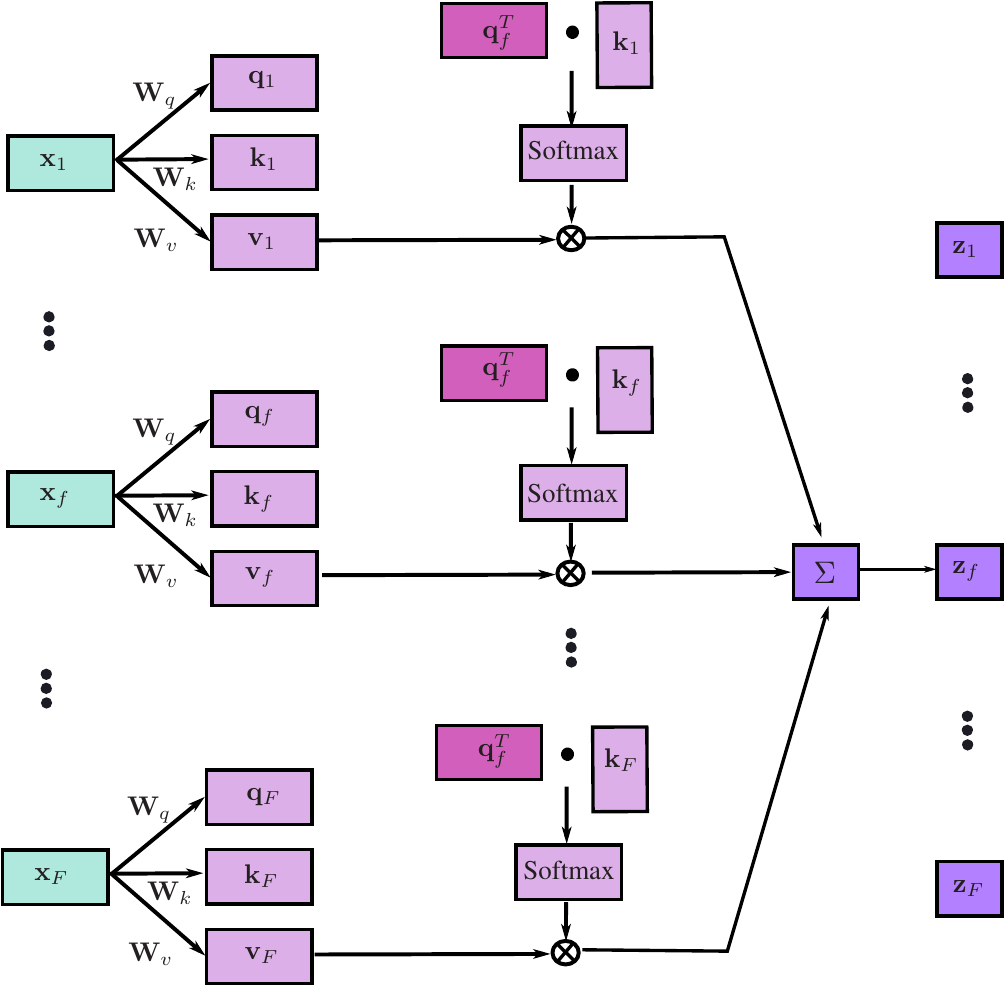}
\caption{An illustration of basic attention mechanism to generate $\mathbf{z}_j$ and same mechanism can be applied to generate $\mathbf{Z}$.}%, $f_{sm}(.)$ represents the softmax function defined in \eqref{eq4}. 
\label{Attention}   
\end{figure}
\subsubsection{Fundamental basics of the attention operation}
%The attention mechanism has been broadly applied in a variety of tasks such as natural language processing and computer vision. 
An example of the attention block is illustrated in Fig.~\ref{Attention}, which takes a matrix $\mathbf{X} = [\mathbf{x}_1,...,\mathbf{x}_{F}] \in \mathbb{R}^{N \times F}$ as the input, generating the output matrix $\mathbf{Z} = [\mathbf{z}_1,...,\mathbf{z}_{F}] \in \mathbb{R}^{N \times F}$. Initially, $\mathbf{X}$ are multiplied by three matrices \footnote{The $\mathbf{W}_v$ matrix may not necessarily have the same size as $\mathbf{W}_q$ and $\mathbf{W}_k$, in this work $N'= N$ for simplicity.}, namely, the query matrix $\mathbf{W}_q \in \mathbb{R} ^{N \times N}$, the key matrix $\mathbf{W}_k \in \mathbb{R} ^{N \times N}$ and the value matrix $\mathbf{W}_v \in \mathbb{R}^{N' \times N}$.  The multiplication operations yield three matrices $\mathbf{Q}, \mathbf{K}, \mathbf{V} \in \mathbb{R}^{N \times N}$, specifically,
\begin{equation}
    \mathbf{Q} = \mathbf{W}_q\hspace{1pt}\mathbf{X}, \hspace{2pt}
    \mathbf{K} = \mathbf{W}_k\hspace{1pt}\mathbf{X}, \hspace{2pt}
    \mathbf{V} = \mathbf{W}_v\hspace{1pt}\mathbf{X}.
    \label{eq2}
\end{equation}
The elements of these three matrices act as hyperparameters that can be fine-tuned during the training process. 
The second step is to calculate the pairwise correlations between all columns of matrices $\mathbf{Q}$ and $\mathbf{K}$, resulting in a new matrix $\mathbf{A} \in \mathbb{R}^{F \times F}$, specifically, 
\begin{equation}
    \mathbf{A} =\frac{1}{\sqrt{N}}\hspace{1pt}\mathbf{Q}^T\mathbf{K}.
\end{equation}
We then apply the \textit{softmax} operation to normalize $\mathbf{A}$ and obtain another matrix $\tilde{\mathbf{A}} \in \mathbb{R}^{F \times F}$. Each element $\tilde{\mathbf{A}}_{i,j}$ is positive and the sum of all the elements in each column is equal to $1$. Specifically, $\tilde{\mathbf{A}}_{i,j}$ is calculated as
\begin{equation}
    \tilde{\mathbf{A}}_{i,j} = \frac{\exp{\mathbf{A}_{i,j}}}{\sum_k\exp{\mathbf{A}_{i,k}}}.
    \label{eq4}
\end{equation}
Finally, the output matrix $\mathbf{Z}$ is calculated as
\begin{equation}
  \mathbf{Z} = \mathbf{V}\hspace{1pt}\tilde{\mathbf{A}},
  \label{eq5}
\end{equation}
where each column of $\mathbf{Z}$ represents a weighted sum, and the weights are determined by the corresponding column in $\tilde{\mathbf{A}}$. 

In addition to the fundamental attention operation, we further introduce the \textit{multi-head attention} mechanism that can improve model capabilities. This mechanism employs a total of $\mathcal{P}$ attention heads, each associated with sets of query matrices ($\mathbf{W}_q^1, ..., \mathbf{W}_q^\mathcal{P}$), key matrices ($\mathbf{W}_k^1, ..., \mathbf{W}_k^\mathcal{P}$), and value matrices ($\mathbf{W}_v^1, ..., \mathbf{W}_v^\mathcal{P}$). The multi-head attention mechanism operates in $\mathcal{P}$ steps. In the initial step, the matrices $\mathbf{W}_q^1, \mathbf{W}_k^1, \mathbf{W}_v^1$ are applied to the input matrix $\mathbf{X}$ following equations \eqref{eq2}-\eqref{eq5}, resulting in the output $\mathbf{Z}_1 \in \mathbb{R}^{N \times F}$. This process is then repeated $\mathcal{P}-1$ times, generating additional output matrices $\mathbf{Z}_2, ..., \mathbf{Z}_P \in \mathbb{R}^{N \times F}$. Finally, we concatenate all output matrices obtained from each step, formulating a matrix $\mathbf{Z}_{tl} \in \mathbb{R}^{N \times \mathcal{P} F}$. The final output matrix $\mathbf{Z}' \in \mathbb{R}^{N \times F}$ can then be expressed as 
\begin{equation}
    \mathbf{Z}' = \mathbf{Z}_{tl}\hspace{1pt}\mathbf{W}_O,
    \label{Mul}
\end{equation}
where $\mathbf{W}_O \in \mathbb{R}^{\mathcal{P} F \times F}$ is another hyperparameter matrix. 
\subsubsection{Positioning encoding}
It is important to note that the attention mechanism neglects the inherent sequence order of the input vectors in $\mathbf{X}$. Consequently, when employing such a mechanism, particularly for tasks dependent on the order of vector arrangement, it is imperative to apply a \textit{positioning encoding} technique to incorporate and preserve this sequential information. The idea of positioning encoding is to add another fixed matrix $\mathbf{X}_k \in \mathbb{R}^{N \times F}$ to $\mathbf{X}$ \cite{Attention_google}, a standardized positioning encoding matrix $\mathbf{X}_k$ is
\begin{equation}
\begin{aligned}
    \mathbf{X}_k(x, y) &= \sin\left(\frac{x}{{10000^{y/N}}}\right), \textrm{for odd $y$}; \\
\mathbf{X}_k(x, y) &= \cos\left(\frac{x}{{10000^{(y-1)/N}}}\right), \textrm{for even $y$}.
\end{aligned}
\label{PosE}
\end{equation}

\subsubsection{Residual mechanism, Layer normalization and position-wise FCNN }
After collecting the matrix $\mathbf{Z}'$, we add the input matrix $\mathbf{X}$ to $\mathbf{Z}'$ to obtain the matrix $\tilde{\mathbf{Z}} \in \mathbb{R}^{N \times F}$. We apply the \textit{residual mechanism} since it preserves the original information of the input matrix. The matrix $\tilde{\mathbf{Z}}$ is then fed to a \textit{layer normalization block}, which first vectorizes $\tilde{\mathbf{Z}} $ into a vector $\tilde{\mathbf{z}} \in \mathbb{R}^{NF}$. Subsequently, each element $\tilde{z}_i$ in $\tilde{\mathbf{z}}$ is scaled to derive a new vector $\hat{\mathbf{z}} \in \mathbb{R}^{NF}$ as in \cite{Attention_google}, specifically,
\begin{equation}
     \hat{z}_i = \gamma\frac{\tilde{z}_i - \mu}{\sigma} + \beta, \\
     %\mu &= \frac{1}{M_1M_2}\sum_k z_k, \\
     %\sigma^2 &= \frac{1}{M_1M_2} \sum_k (z_k - \mu)^2,
\label{L7}
\end{equation}
where $\mu$ and $\sigma^2$ represent the mean and variance of vector $\tilde{\mathbf{z}}$. The parameters $\gamma$ and $\beta$ denote the amplitude scaling and the bias, respectively. By default,  $\gamma = 1$ and $\beta = 0$, although these parameters can be adjusted as learning hyperparameters. We then reformulate $\hat{\mathbf{z}}$ into a matrix $\hat{\mathbf{Z}} \in \mathbb{R}^{N \times F}$. To enhance the capacity to capture nonlinear relationships, we feed the output matrix $\hat{\mathbf{Z}}$ into a pointwise FCNN to get $\hat{\mathbf{Z}}' \in \mathbb{R}^{N \times F}$ \cite{Attention_google}, specifically, 
\begin{equation}
    \hat{\mathbf{Z}}' = \mathbf{W}_2\hspace{1pt}f_{\text{Relu}}(\mathbf{W}_1\hspace{1pt}\hat{\mathbf{Z}}+\mathbf{B}_1)+\mathbf{B}_2,
\end{equation}
where $f_{\text{Relu}}(.)$ represents the  rectifier activation function, and $\mathbf{W}_1, \mathbf{W}_2, \mathbf{B}_1, \mathbf{B}_2$ are hyperparameter matrices, and the bias matrices $\mathbf{B}_1, \mathbf{B}_2$ are optional. After collecting $\hat{\mathbf{Z}}'$, we apply the same residual mechanism and layer normalization to derive $\breve{\mathbf{Z}} \in \mathbb{R}^{N \times F}$. Finally, $\breve{\mathbf{Z}}$ is vectorized and fed into another FCNN. Such an operation can also help to match the vector sizes for possible subsequent blocks. 

\subsection{Data cleaning and signal processing}
\label{DC}
The collection of UL SRS channel measurements in a commercial 5G NR BS builds limitations when retrieving data-intense structures such as SRS channel measurement samples. The vast amounts of SRS data generated at milliseconds level are normally enclosed within the baseband entity of a BS and primarily intended for internal processing, whereas external access to these data may be compromised by hardware and software restrictions. To mitigate these challenges, it is essential to equip our pipeline with the ability to discern the validity of the input data. Accordingly, we introduce a \textit{data-cleaning} block to pre-process the measurement data. Its primary objective is to determine whether the raw transfer function is valid or invalid. A raw transfer function is labeled invalid if it satisfies any of the following criteria:
\begin{itemize}
    \item Insufficient CSI in the beam or frequency domain: the number of non-zero elements in $\mathbf{H}_t$ is lower than a given threshold.
    \item Update failure: the values of all subcarriers or all beams remain the same.
%    \item The phase along all subcarriers of one beam loses coherency.
%    \item The Frobenius norm of the CTF at the current snapshot is significantly smaller than that of neighbouring snapshots.
\end{itemize}
%The labels are used to train the ML-based data-cleaning framework, comprising multiple attention blocks and utilizing the sigmoid function at the final layers to predict probabilities for each label. We denote all hyperparameters as $\boldsymbol{\theta}$ and $\hat{p}_i = f_{dc}(\theta, \mathbf{H}_i)$ as the probability of predicting label 0 of the $i$-th input matrix $\mathbf{H}_i$. During the training process, the cross-entropy function is chosen as the objective criterion and the total loss $\Psi_1$ is defined as
%\begin{equation}
  %  \Psi_1 = -\sum_{i \in \Omega_{tr}} \kappa_i\hspace{1pt}\log(\hat{p}_i) + (1-\kappa_i)\hspace{1pt}\log(1-\hat{p}_i),
%\end{equation}
%where $\Omega_{tr}$ denotes the training dataset while $\kappa_i$ is the ground-truth label of the $i$-th input. 
After filtering out all invalid data, the next step is to process the raw CTF to generate impulse response beam matrices. To suppress the side lobes, we apply Hann windowing across all rows of the matrix $\mathbf{H}_t$ to obtain matrix $\hat{\mathbf{H}}_t \in \mathbb{C}^{N \times F}$. The  $F$-length Hann window in the frequency domain is given by
\begin{equation}
    w[f] = \sin^2\left(\frac{\pi f}{F}\right), \quad f= 0, \ldots, F-1.
\end{equation}
After the windowing operation, the impulse response beam matrix $\mathbf{G}_t$ is produced by performing the inverse discrete Fourier transform along each row of $\hat{\mathbf{H}}_t$. Given the potential difficulty in achieving a stable phase for $\mathbf{G}_t $, here we opt to use its amplitude $|\mathbf{G}_t|$ as the training fingerprint, although this means throwing away potentially useful information.

\subsection{Single-snapshot localization}
\label{SsLa}
As illustrated in Fig.~\ref{Pipeline}, the architecture comprises multiple attention-aided blocks, followed by an output layer that has three alternatives corresponding to three loss functions, namely the Mean Square Error (MSE), Negative Log-Likelihood (NLL), and Regression-by-classification loss functions. We use $\mathbf{p}_i = [p_{x,i}, p_{y,i}]^T$ to represent the $2$-D ground truth of the moving UE at the $i$-th position. Notably our approach can be readily adapted for $3$-D localization. 
\subsubsection{Alternative 1: MSE loss function}
This approach directly estimates the UE locations by setting a $2$-D regression head at the output layer of the last attention block. Let $f_{\text{MSE}}(.)$ denote the overall function and vector $\boldsymbol{\theta}_{2}$ all hyperparameters, $\hat{\mathbf{p}}_i = [\hat{p}_{x,i}, \hat{p}_{y,i}]^T$ the estimated $i$-th UE locations generated by $f_{\text{MSE}}(\boldsymbol{\theta}_{2}, |\mathbf{G}_t|)$, the loss $\Psi_{1}$ can be expressed as
\begin{equation}
    \Psi_1 = \frac{1}{N_{tr}}\sum_{i \in \Omega_{tr}'}||\mathbf{p}-\hat {\mathbf{p}}||^2_F,
    \label{MSE}
\end{equation}
where $\Omega_{tr}'$ and $N_{tr}$ denote the training set and the number of training samples, respectively, and $||.||_F$ denotes the Frobenius matrix norm. \vspace{-0pt} 
\subsubsection{Alternative 2: NLL loss function}
Unlike the first approach, this method employs the NLL criterion, which models the estimated UE position as a multivariate Gaussian distribution defined by its mean $\breve{\mathbf{p}} = [\breve{p}_{x_i}, \breve{p}_{y_i}]^T$ and variance $\breve{\boldsymbol{\sigma}}_i^2 = [\breve{\sigma}^2_{x_i}, \breve{\sigma}^2_{y_i}]^T$. Consequently, a $4$-dimensional regression head is required at the output layer. Similar to \cite{Guoda1}, the NLL loss $\Psi_2$ is expressed as \vspace{-1pt}
\begin{equation}
  \Psi_2 = \frac{1}{2N_{tr}}\sum_{i \in \Omega_{tr}'}\Big(\frac{\log \breve{\sigma}^2_{x_i}\hspace{1pt}\breve{\sigma}^2_{y_i}}{2} +  \frac{(p_{x_i} - \breve{p}_{x_i})^2}{2\breve{\sigma}^2_{x_i}} +\frac{(p_{y_i} - \breve{p}_{y_i})^2}{2\breve{\sigma}^2_{y_i}}\Big).
  \label{NLL}
\end{equation} 
\subsubsection{Alternative 3: Regression-by-Classification (RbC)}
The core of this approach \cite{Adabin, Xiong} lies in converting a regression task to a classification task. This is achieved by first defining a feasible range for the target parameter and then dividing this range into discrete bins. For the localization task, the lower and upper bounds of the UE x-coordinates are denoted as $B_{lw,x}$ and $B_{up,x}$, respectively. Similarly, $B_{lw,y}$ and $B_{up,y}$ represent the bounds for the $y$-coordinates. To accomplish this discretization, we divide the  $x$-coordinate range into $L_x$ equally sized bins. The $y$-coordinate range is divided into $L_y$ bins in a similar fashion. For each bin, we denote $\bar{l}_{x,k}$ and $\bar{l}_{y,k}$ as the lower endpoint values of the $k$-th interval for the $x$- and $y$-coordinates, respectively. %We then define vectors $\boldsymbol{\xi}_x \in \mathbb{R}^{L_x}$ and $\boldsymbol{\xi}_y \in \mathbb{R}^{L_y}$ to collect all $\bar{l}_{x,i}$ and $\bar{l}_{y,i}$.

Unlike the NLL method, RbC does not inherently model the output probability as a Gaussian distribution. Instead, it estimates the probability and bias values of each bin for both the $x$- and $y$-coordinates. The bias value can be used to reduce the quantization error. To this end, in total $4$ vectors are generated: the probability vectors $\mathbf{q}_x \in \mathbb{R}^{L_x}$ and $\mathbf{q}_y \in \mathbb{R}^{L_y}$, as well as the deviation vectors $\mathbf{d}_x \in \mathbb{R}^{L_x}$ and $\mathbf{d}_y \in \mathbb{R}^{L_y}$. It is crucial to apply a \textit{softmax} operation as shown in \eqref{eq4} when generating $\mathbf{q}_x$ and $\mathbf{q}_y$ to ensure that the elements within each vector sum to~1. One special case for deviation vectors is when all $L_x$ elements in $\mathbf{d}_x$ have the same value, and the same for $\mathbf{d}_y$. In other words, a uniform shift is applied to the probability density function, which also aids in the reduction of the output vector dimensions. We denote $q_{x,k}$ and $d_{x,k}$ as the $k$-th elements of $\mathbf{q}_x$ and $\mathbf{d}_x$, similarly for $q_{y,k}$ and $d_{y,k}$. Inspired by \cite{Adabin}, the $\eta$-norm loss $\Psi_3^\eta$ is formulated as \vspace{-5pt}
\begin{align}
    &\Psi_3^\eta = \frac{1}{2N_{tr}}\sum_{i \in \Omega_{tr}'}\Big(||\sum_{k=1}^{L_x}q_{x,k,i}\bar{l}_{x,k,i}-p_{x,k,i}+d_{x,k,i}||^\eta 
    \nonumber\\&+||\sum_{j=1}^{L_y}q_{y,j,i}\bar{l}_{y,j,i}-p_{y,j,i}+d_{y,j,i}||^\eta + \gamma_1||\mathbf{d}_x|| + \gamma_2 ||\mathbf{d}_y|| \Big).    
    \vspace{-5pt}
    \label{RbC}
\end{align}
Here, $\eta$ is usually chosen as $\eta = 1$ or $\eta = 2$, which corresponds to the \textit{Taxicab} and \textit{Euclidean} norms,  respectively. Two penalty terms, $\gamma_1||\mathbf{d}_x||$ and $\gamma_2 ||\mathbf{d}_y||$, are added to the cost function. The estimated coordinate $\hat{\mathbf{p}}_{i}^{\text{RbC}} = [\hat{p}_{x,i}^{\text{RbC}},\hat{p}_{y,i}^{\text{RbC}}] \in \mathbb{R}^2$ is then given by 
\begin{align}
    \hat{p}_{x,i}^{\text{RbC}} &= \sum_k q_{x,k,i}\bar{l}_{x,k,i} + d_{x,k,i}, \nonumber\\
    \hat{p}_{y,i}^{\text{RbC}} &= \sum_k q_{y,k,i}\bar{l}_{y,k,i} + d_{y,k,i}. 
\end{align}
\subsubsection{Comparison between different uncertainty estimates}
Our previous work \cite{Guoda1} used the NLL score in the test data set to assess the effectiveness of uncertainty estimation. However, applying the same criterion to evaluate the RbC method presents challenges because of the non-Gaussian nature of its output. To address this challenge, another criterion named Area Under the Sparsification Error (AUSE) \cite{AUSE} is used.
Sparsification is a way to assess the quality of uncertainty estimates. It works by progressively discarding
fractions of the predictions that the model is most uncertain about and verifying whether this corresponds to a proportional decrease in the remaining average endpoint error. To calculate AUSE, the first step is to compute the discrete entropy $u_H$ based on the predicted probability. In the following discussion, we illustrate this process using the predicted $\mathbf{q}_{x,i}$ vector for the $x$-coordinate as an example, noting that the result can be readily extended to the $y$-coordinate. The entropy $u_{H,x,i}$ for $\mathbf{q}_{x,i}$ is given by \cite{Xiong}
\begin{equation}  
   u_{H,x,i}(\mathbf{q}_x) = -\sum_{k=1}^{L_x} q_{x,k,i}\hspace{1pt}\log q_{x,k,i}.
   \label{entropy}
\end{equation}
To enable a fair comparison between the NLL and RbC methods, we need to discretize the predicted Gaussian distributions determined by $\breve{\mathbf{p}}$ and $\breve{\sigma}_i^2$. To this end, the $x$-axis is segmented into $L_x$ bins. As detailed in \cite{DGaussian}, the value for the $k$-th bin of the discretized function, denoted $\breve{p}_{x,k}$, is calculated as
\begin{equation}
    \breve{p}_{x,k}^{ds} = \frac{\frac{1}{ \breve{\sigma}_k}\hspace{1pt}\exp(-\frac{(\breve{p}_{x,k}-\bar{l}_k)^2}{2\breve{\sigma}_k^2})}{\sum_j \frac{1}{ \breve{\sigma}_j}\hspace{1pt}\exp(-\frac{(\breve{p}_{x,j}-\bar{l}_j)^2}{2\breve{\sigma}_j^2})}.
    \label{DGaussain}
\end{equation}
We now organize the discrete entropies for the $N_{ts}$ testing samples calculated from \eqref{entropy} in descending order to form the vector $\mathbf{u}_{H,x} \in \mathbb{R}^{L_x}$. Similarly, we calculate the absolute errors between the estimated values $\hat{p}_{x,i}^{\text{Ws}}$ and the ground truth $p_x$ for all testing samples, arranging these errors in descending order to create the vector $\boldsymbol{\xi}_x \in \mathbb{R}^{L_x}$. Let $\xi_{\text{max}}$ be the maximum absolute error. We scale all elements in $\mathbf{u}_{H,x}$ by a factor such that the first element of the resulting vector $\hat{\mathbf{u}}_{H,x}$ equals $\xi_{\text{max}}$. 

Next, we define a \textit{sparsification} function $s(\varphi)$, which is calculated by removing the initial $\varphi$-fraction of samples from $\hat{\mathbf{u}}_{H,x}$ and averaging the remaining data, with $\varphi$ ranging from 0 to 1. A similar process is applied to $\boldsymbol{\xi}_x$, which yields the oracle function $g(\varphi)$.  Finally, AUSE is calculated as
\begin{equation}
\textrm{AUSE} = \int_{0}^1 |s(\varphi) - g(\varphi)|\hspace{1pt}d\varphi,
\label{EqAUSE}
\end{equation}
which represents the area between the sparsification and the oracle curves. We refer to Fig.~\ref{Spar} as an example, where AUSE equals the area of the shaded region, a smaller area indicating a better uncertainty estimator. %This parameter serves as a comprehensive measure to assess the performance of the uncertainty estimation method. 
%Specifically, the first alternative applies MSE criteria, which does not estimate uncertainty. The second one applies the NLL criteria, which treats the output as Gaussian distributed and predicts both the mean position and the variances. The third alternative uses Wasser can not only predict the position, but also its variances  
\subsection{Kalman-Filter-based trajectory smoothing}
To further improve the localization accuracy, we exploit the temporal correlation between successive positions by applying a Kalman filter as a straightforward method for trajectory smoothing. For more detailed information see \cite{Kay}. We define a vector $\boldsymbol{\xi}_t \in \mathbb{R}^4 = \left[p_{x,t}, v_{x,t}, p_{y,t}, v_{y,t}\right]^T$ to represent the UE position and velocity at time $t$, where $v_{x,t}$ and $v_{y,t}$ denote the speed in the $x$ and $y$-directions, respectively. The state-space model for the UE is given by
\begin{equation}
    \boldsymbol{\xi}_t =  \mathbf{F}\hspace{1pt}\boldsymbol{\xi}_{t-1} + \boldsymbol{\lambda}_t,
    \label{KFfirst}
\end{equation}
where $\mathbf{F} \in \mathbb{R}^{4 \times 4}$ denotes the state-transition matrix, while $ \boldsymbol{\lambda}_t \in \mathbb{R}^4$ the additive noise. Specifically, 
\begin{equation}
    \mathbf{F} = \begin{bmatrix}
1 & 0 & \Delta_t & 0\\
0 & 1 & 0 &  \Delta_t \\
0 & 0 & 1 & 0\\
0 & 0 & 0 & 1
\end{bmatrix},	
\end{equation}
where $\Delta_t$ denotes the time differences between snapshots. 
We then define $\boldsymbol{\Xi}_t \in \mathbb{R}^{4 \times 4}$ as the covariance matrix of $\boldsymbol{\xi}_t$. The relationship between $\boldsymbol{\Xi}_t$ and $\boldsymbol{\Xi}_{t-1}$ can be written as 
\begin{equation}
    \boldsymbol{\Xi}_t = \mathbf{F}\boldsymbol{\Xi}_{t-1}\mathbf{F}^T + \boldsymbol{\Lambda},
    \label{XiGamma}
\end{equation}
where $\boldsymbol{\Lambda} \in \mathbb{R}^{4 \times 4}$ is the covariance matrix of the noise vector~$\boldsymbol{\lambda_t}$.  
We further denote $\breve{\mathbf{p}}_t \in \mathbb{R}^2 = \left[\breve{p}_{t,x}, \breve{p}_{t,y}\right]$ as the predicted UE position and express the observation model as 
\begin{equation}
    \breve{\mathbf{p}}_t = \boldsymbol{\Phi}_t\hspace{1pt}\boldsymbol{\xi}_t + \boldsymbol{\zeta},
\end{equation}
where $\boldsymbol{\zeta} \in \mathbb{R}^2$ represents  observation noise and $\boldsymbol{\Phi}_t = \begin{bmatrix}
1 & 0 & 0 & 0\\
0 & 1 & 0 & 0 \\ 
\end{bmatrix} $. Given the error signal $\mathbf{e}_t = \hat{\mathbf{p}}_t -\breve{\mathbf{p}}_t$, the state vector $\boldsymbol{\xi}_t^{+}$ is updated as
\begin{equation}
    \boldsymbol{\xi}_{t}^{+} =  \boldsymbol{\xi}_{t} + \mathbf{\Gamma}_t\hspace{1pt}\mathbf{e}_t.
    \label{update}
\end{equation}
In \eqref{update}, $\mathbf{\Gamma}_t$ represents the Kalman gain matrix, which balances the predictions from the state-space model and the ML-based pipeline, specifically,
\begin{equation}
    \mathbf{\Gamma}_t = \boldsymbol{\Xi}_t\hspace{1pt}\boldsymbol{\Phi}_t^T\left[\boldsymbol{\Phi}_t\hspace{1pt}\boldsymbol{\Xi}_t\boldsymbol{\Phi}_t^T + \mathbf{R}\right]^{-1},
    \label{GammaR}
\end{equation}
where $\mathbf{R}$ is the covariance matrix of $\boldsymbol{\zeta}$. After computing $\boldsymbol{\Gamma}_t$, the covariance matrix $\boldsymbol{\Xi}_t$ is updated using
\begin{equation}
    \boldsymbol{\Xi}_t^+ = (\mathbf{I}-\boldsymbol{\Gamma}_t\hspace{1pt}\boldsymbol{\Phi}_t)\hspace{1pt}\boldsymbol{\Xi}_t,
    \label{FinalKF}
\end{equation}
where $\mathbf{I}$ denotes the identity matrix. By applying the process outlined by \eqref{KFfirst}-\eqref{FinalKF}, we can significantly mitigate the impact of prediction outliers, as will be further illustrated in Section \ref{Section5}. 
\section{Outdoor 5G NR measurement campaign}
\label{Section4}
To evaluate our localization pipeline, an outdoor vehicular measurement campaign was conducted at a parking lot outside of the Ericsson office in Lund, Sweden. Photos of the test vehicle, the BS antenna, the UE as well as the measurement areas are presented in Fig.~\ref{Measurement}.  %{-7pt}

\subsection{Introduction to the measurement campaign}
\begin{figure}[b!]
	\centering  
	%{-0.30cm} 
    \includegraphics[width=1.15\linewidth]{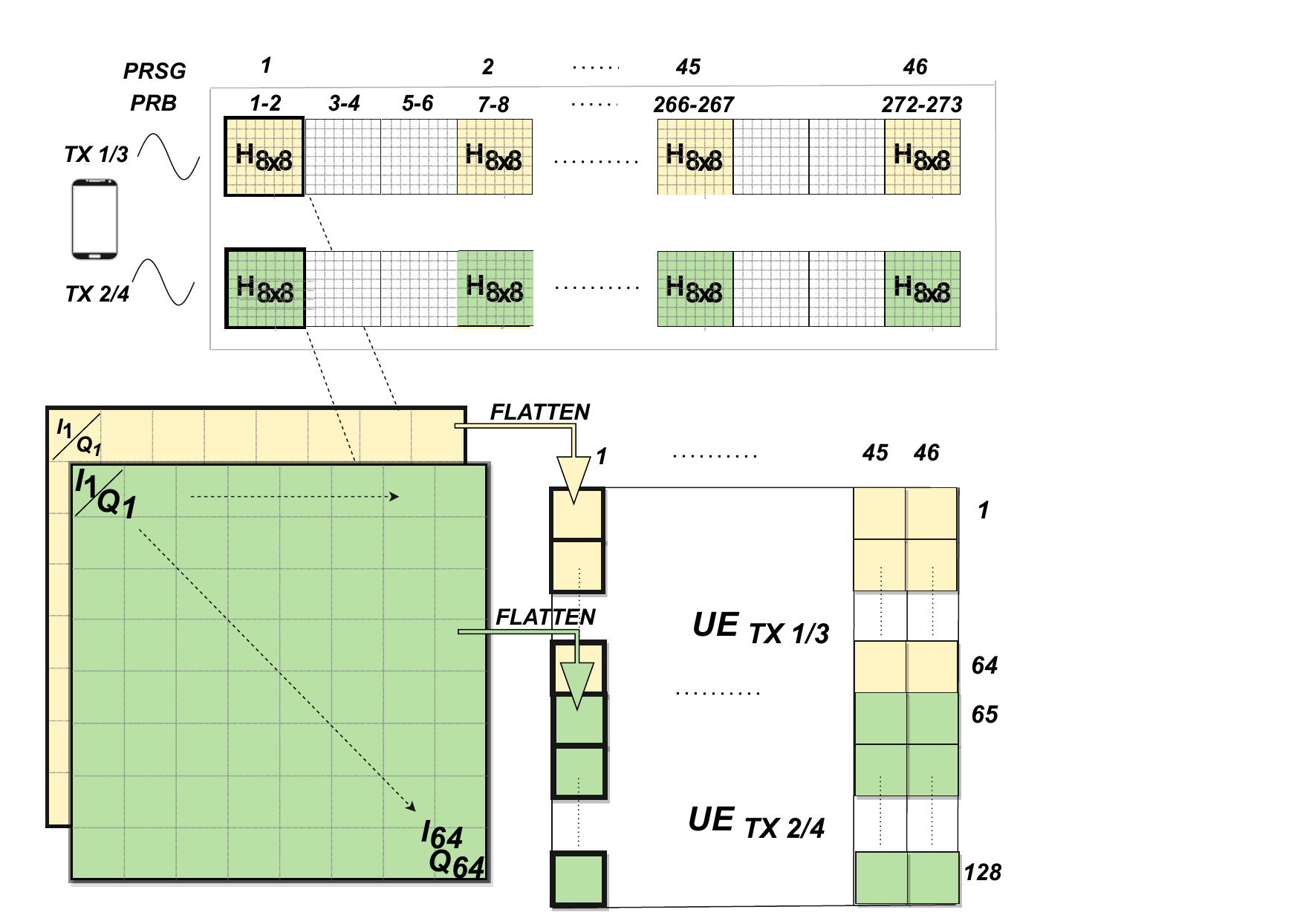}%NNCostupdate\
	\caption{SRS data collection and CTF generation.}
	\label{SRS-structure} 
\end{figure}
%\vspace{-45pt}
\begin{figure*}[h!]
\centering \centerline{\includegraphics[width=2.0\columnwidth,height = 5.0cm]{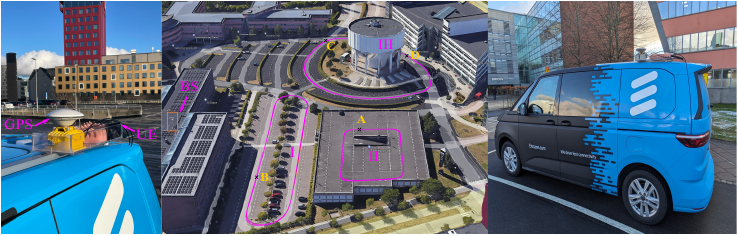}}      
\caption*{\hspace{-30pt}(a) GPS and UE\hspace{90pt} (b) Measurement Scenario  \hspace{110pt} (c) Measurement van} \vspace{-0pt}
	\caption{The 5G NR base station was equipped with an antenna integrated radio with 64 transmitters and receivers, placed on top of a 20 m high building. In this measurement campaign, a vehicle moves along three pre-defined routes: \textbf{I} A route on a $10$ 
 meter-high garage for LOS measurements. \textbf{II}: A ground-level route for NLoS measurements below the building of the BS. \textbf{III}: A ground-level route for combined LoS and NLoS measurements.}
\label{Measurement} 
\end{figure*}
During the measurement campaign, the test vehicle carried a GNSS receiver, and a commercial UE, see Fig.~\ref{Measurement}(a). Centimeter-level ground truth positioning accuracy was achieved using a Swift Duro high-performance GNSS receiver with Real Time Kinematics (RTK) technology, GNSS multi-band and multi-constellation support. To ensure that the UE remained in connected state, it simultaneously downloaded data at a 750 Mbit/s rate enabling continuous SRS UL transmission. The UL SRS pilot signals were received and processed by a commercial Ericsson 5G BS operating in mid-band at $3.85$ GHz center frequency. The BS was compliant to the 5G NR 3GPP standard 38.104 Rel15 \cite{3GPP-38104} and equipped with a TDD antenna integrated radio with 64 transmitters/receivers (TX/RX) consisting of $32$ dual-polarized antennas covering a $120$ degree sector. As for digital beam forming, 64 TX/RX formulate 64 beams in DL/UL respectively. As illustrated in Fig. \ref{SRS-structure}, the SRS channel estimates are reported for 273 PRBs over a 100 MHz bandwidth. Each channel snapshot contains the 273 PRBs for all 64 beams. The PRBs are grouped and averaged in pairs, resulting in $137$ Physical Resource Blocks Sub Groups (PRSG). Down sampling was done so that every third PRSG was further used generating 46 PRSGs in total. The UE was equipped with 4 antenna ports, i.e. 4 UE layers, sounding SRS pilots. % while alternating 2 antenna ports at the time forming $4$ layers. 
Due to the capacity of our data-streaming system, the BS recorded the channel responses of $2$ UE antenna ports which formulate two channel transfer function matrices $\mathbf{H}_1, \mathbf{H}_2 \in \mathbb{C}^{N \times F}$. We define a matrix $\mathbf{H}' \in \mathbb{C}^{2N \times F}$ to collect those two matrices, specifically, $\mathbf{H}' = [\mathbf{H}_1, \mathbf{H}_2]$ ($N = 64, F = 46$).  %After initial post-processing, the final CTF dimension for a single channel snapshot is 128 x 46 (5888) standing for two UE layers collected over 46 PRSGs. %It is noteworthy that only the amplitude information from I/Q samples was utilized as input to the ML model.
As illustrated in Fig. \ref{Measurement}, our measurement campaign comprises three distinct scenarios: LoS, NLoS, and a mixed scenario. In all scenarios, the velocity of the vehicle is approximately $15$ km/h. The trajectory for each of the three measurement scenarios consists of $5$ laps. In the LoS scenario, the test vehicle drove at an open parking lot, while in the NLoS scenario, the vehicle was driving next to a tall building that obstructed the LoS path. As for the mixed scenario, NLoS conditions occurred when the LoS was blocked by the water tower. For all three measurements, the BS station recorded channel snapshots with $20$ ms periodicity,  resulting in $\mathcal{T}_{1} = 22000$, $\mathcal{T}_{2} = 24603$ and $\mathcal{T}_{3} = 27087$ channel snapshots. We formulate three tensors $\mathcal{A}_{\text{LoS}} \in \mathbb{C}^{\mathcal{T}_{1} \times 2N \times F}$, $\mathcal{A}_{\text{NLoS}} \in \mathbb{C}^{\mathcal{T}_{2} \times 2N \times F}$, $\mathcal{A}_{\text{mix}} \in \mathbb{C}^{\mathcal{T}_{3} \times 2N \times F}$ to collect all snapshots. Those three tensors are normalized by multiplying each with a scalar so that their Euclidean norms equals $\mathcal{T}_i\hspace{1pt}MN$, where $i = 1,2,3$.
\begin{figure*}[t!]
\begin{minipage}[t]{0.5\textwidth}
\hspace{-5pt}
  \includegraphics[width=1.00\linewidth]{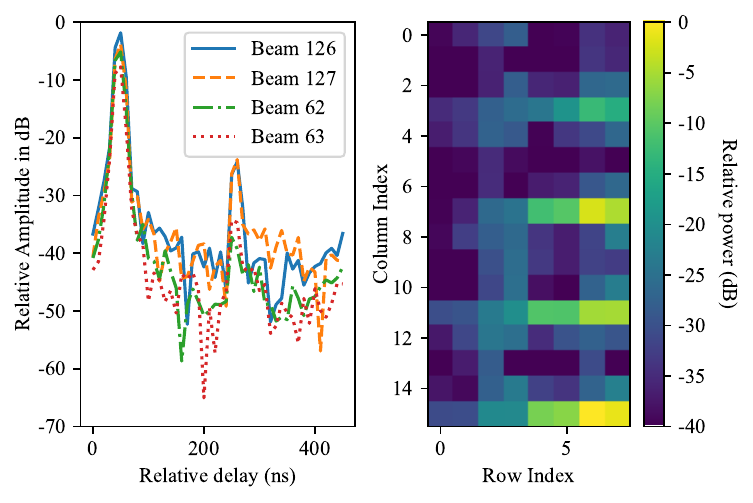}
\captionsetup{labelformat=empty}\addtocounter{figure}{-1}\caption{(a)}
  \label{fig:first}
\end{minipage}% 
\hfill\allowbreak % maximize the horizontal separation
\begin{minipage}[t]{0.5\textwidth}
\hspace{-5pt}
\includegraphics[width=1.00\linewidth]{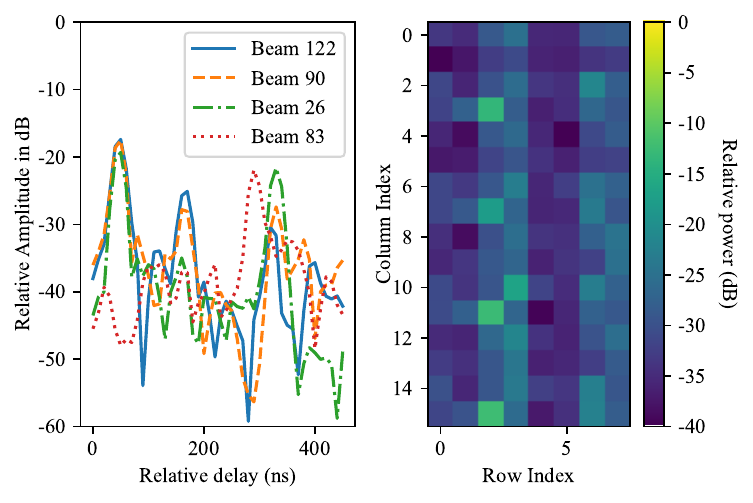}
\captionsetup{labelformat=empty}\addtocounter{figure}{-1}\caption{(b)}
  \label{fig:second}
\end{minipage}%
\hfill\allowbreak
\begin{minipage}[t]{0.5\textwidth}
\hspace{-5pt}
 \includegraphics[width=1\linewidth]{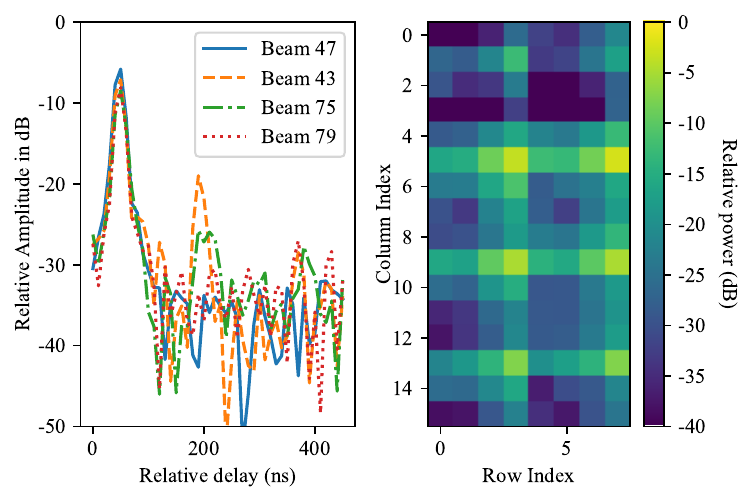}
   \captionsetup{labelformat=empty}\caption{(c)}
   \label{fig:third}
\end{minipage}%
\hfill\allowbreak
\begin{minipage}[t]{0.5\textwidth}
\hspace{-5pt}
 \includegraphics[width=1\linewidth]{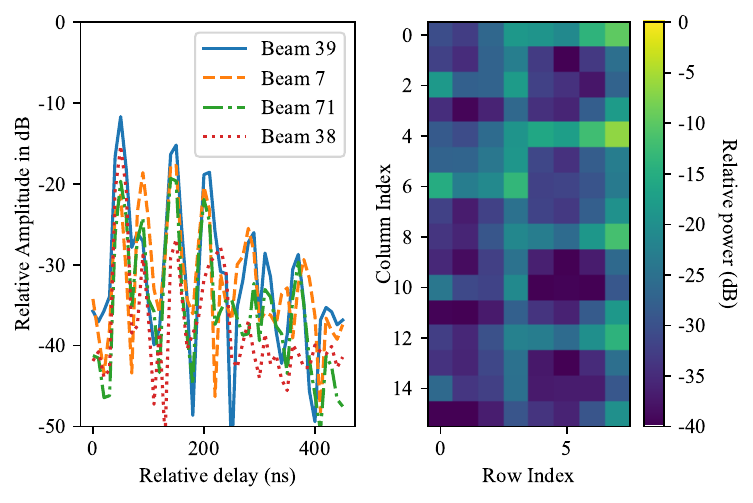}
   \captionsetup{labelformat=empty}\caption{(d)}
   \label{fig:forth}
\end{minipage}
\addtocounter{figure}{-1}\caption{CIR and relative power of all $128$ beams of four locations (a) LoS at point A, (b) NLoS at point B, (c) LoS at point C, (d) NLoS at point D. Beam diagrams are arranged as follows: row $0-3$ and row $4-7$ represent the $32$ horizontal and $32$ vertical-polarized beams respectively for UE layer $1$; row $8-11$ and row $12-15$ represent the co-polarized beams for UE layer $2$. Beam index is $8*(i-1)+j$, where $i$ and $j$ denote the row and column index respectively. We select the first $4$ strongest beam and plot the relative amplitude of CIR.} \vspace{-10pt}
 \label{fig:Mea}
\end{figure*}

\subsection{Measured propagation channel characteristics}
We choose four UE positions (positions A-D, see Fig.~\ref{Measurement}~(b)) from the three measurement scenarios and show representative CIRs in Fig.~\ref{fig:Mea}~(a)-(d). To be specific, Fig.~\ref{fig:Mea} (a) illustrates a typical LoS scenario where a dominant LoS path can be seen from both the CIR and the beam patterns. Few beams exhibit dominant power levels, while others remain comparatively weaker. %Since there exists an overlap between beam $59$ and $58$, their signal strengths of the LoS path are similar. 
Although few NLoS-paths can still be observed, their strengths are much weaker compared to the direct path. This is because the UE is located in an open parking lot, where the reflected signals from other buildings are relatively weak. From the beam power pattern, one can observe the signal strength variations of different BS antenna polarizations and UE transmission layers as well. In contrast, Fig.~\ref{fig:Mea} (b) displays NLoS channel characteristics where the BS captures several reflected paths and there is no path with a dominant power. Thus, the signal strength in Fig.~\ref{fig:Mea} (b) is lower compared to the case in Fig.~\ref{fig:Mea} (a). Fig.~\ref{fig:Mea}~(c) and Fig.~\ref{fig:Mea}~(d) present the measured channels in a mixed scenario, where more local scatters surround the UE. The distance between UE position C and the BS is greater than that of UE position A, resulting in a decrease in the strength of the received LoS signal. Nevertheless, the BS is capable of detecting stronger reflective paths in addition to the LoS path, attributed to reflections from surrounding buildings.  Similarly, in Fig.~\ref{fig:Mea} (d), a rich number of multipath components can be observed in both the CIR and the beam pattern, despite the LoS path being obstructed.         
\section{Results and discussion}
\label{Section5}
In this section, we evaluate our ML-based localization pipeline using the measurements. %Initially, we assess the effectiveness of our data-cleaning approach. 
We initially compare the single-snapshot localization performance for different ML algorithms under different scenarios. Then, we demonstrate the performance gain achieved by smoothing multiple position estimates with a Kalman filter. 

\begin{table}
\centering
\caption{Overview of our ML-based single snapshot localization pipeline}
\vspace{-5pt}%, the numbers in parentheses represent the distances between two training samples
{\begin{center}
\begin{tabular}{ccccccc}
\hline\hline
Item & Network Structures or Parameters  \\
\hline
Input Features & Amplitude of CIRs for all beams\\
Network Output & Estimated position labels or probabilities\\ 
%Middle layer 1 & Layer normalization according to (\ref{L7}) \\
Intermediate block 1 & Residual $2$-Heads Self-attention Network \\ 
Intermediate block 2 & Residual Position-wise FCNNs \\
Intermediate block 3 & $3$ cascaded ordinary FCNNs \\
Time Complexity & $NF^2$\\
\hline\hline
\end{tabular}
\end{center}
}
\label{table: NLL1}
\vspace{-10pt}
\end{table}
%We investigate our data-cleaning approach on received CSI across three distinct scenarios.  

%The classification performance of our approach in three testing scenarios is depicted in Fig. \ref{Class_accu}. The results demonstrate a precision exceeding $98\%$ accuracy across all test cases. It is important to note that the cutting-off threshold is not strictly "black-or-white". Consequently, a few misclassified samples may occur when the number of non-zero elements hovers around a borderline. Despite these occasional misclassifications, our method is deemed effective for the data-cleaning task.  

\subsection{Single snapshot localization}
\begin{figure}[t!]
  \centering
\begin{subfigure}{}
  \centering
  % include first image
  \includegraphics[width=0.9\linewidth]{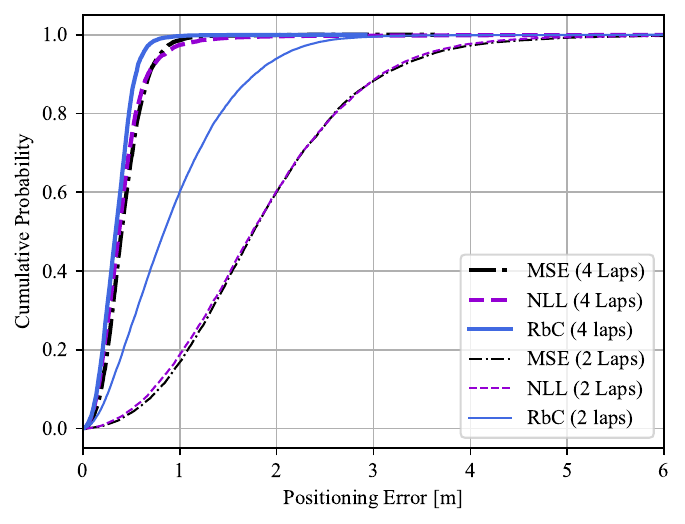}  
  \captionsetup{labelformat=empty}
  \addtocounter{figure}{-1} \vspace{-5pt}
  \caption{(a)} \vspace{-5pt}%Positioning error cumulative distribution function when adjacent samples are separated with a distance $\frac{1}{8}\lambda$.}
  \label{LoSAcc}
\end{subfigure}
\begin{subfigure}
  \centering
  % include second image
  \includegraphics[width=0.9\linewidth]{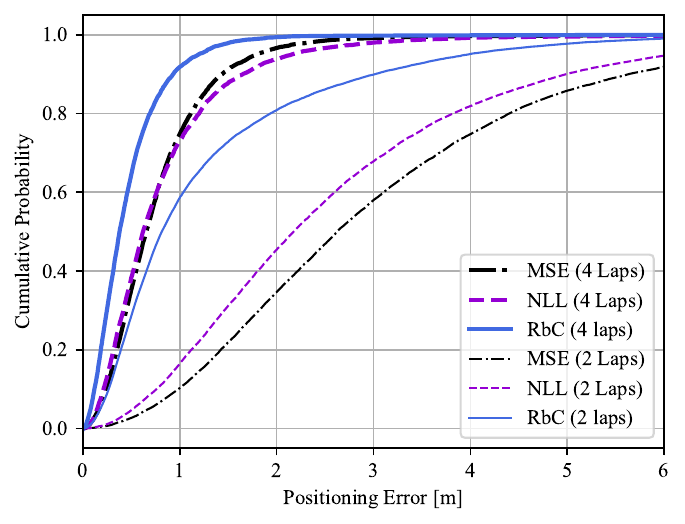}  
  \captionsetup{labelformat=empty}
  \addtocounter{figure}{-1}\vspace{-5pt}
  \caption{(b)} \vspace{-5pt}% Positioning error cumulative distribution function when adjacent samples are separated with a distance $\frac{3}{4}\lambda$.}
  \label{NLoSAcc}
\end{subfigure}
\begin{subfigure}
  \centering
  % include second image
  \includegraphics[width=0.9\linewidth]{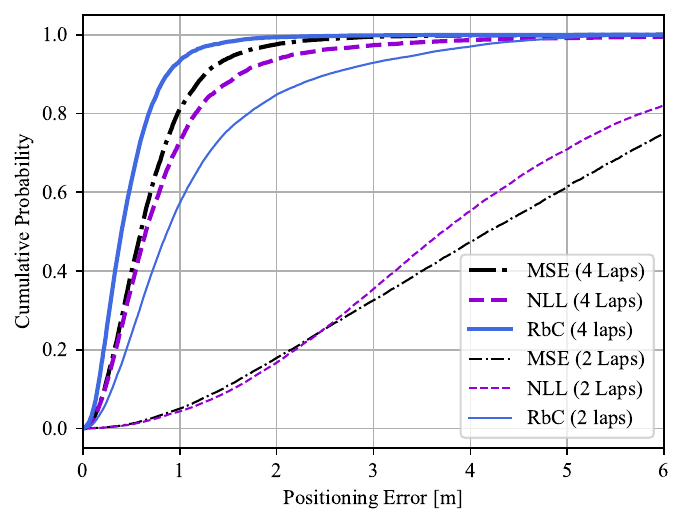}  
  \captionsetup{labelformat=empty}
  \addtocounter{figure}{-1}\vspace{-5pt}
  \caption{(c)} \vspace{-5pt}% Positioning error cumulative distribution function when adjacent samples are separated with a distance $\frac{3}{4}\lambda$.}
  \label{MixAcc}
\end{subfigure}
\caption{Positioning errors of different training densities in the three scenarios: (a) LoS, (b) NLoS, (c) Mixed.}
\label{Single_SP_PoS} 
\vspace{-20pt}
\end{figure}

\begin{figure}[t!]
  \centering
\begin{subfigure}{}
  \centering
  % include first image
  \includegraphics[width=0.9\linewidth]{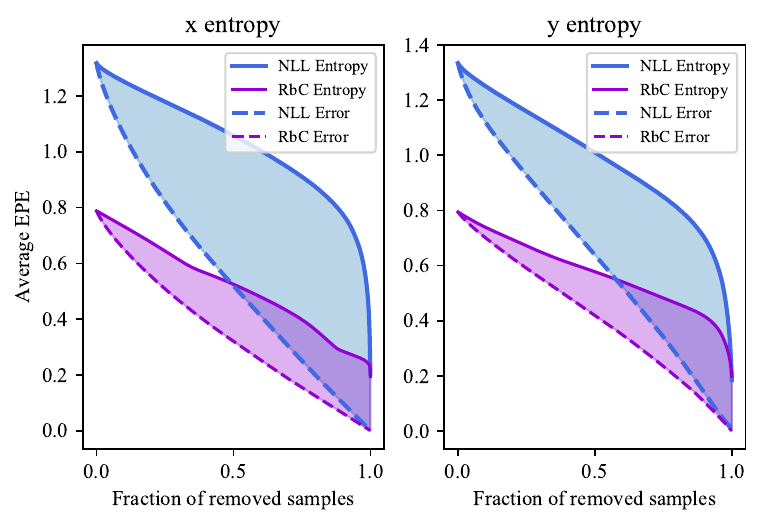}  
  \captionsetup{labelformat=empty}
  \addtocounter{figure}{-1} \vspace{-5pt}
  \caption{(a)} \vspace{-5pt}%Positioning error cumulative distribution function when adjacent samples are separated with a distance $\frac{1}{8}\lambda$.}
  \label{LoSAcc}
\end{subfigure}
\begin{subfigure}
  \centering
  % include second image
  \includegraphics[width=0.9\linewidth]{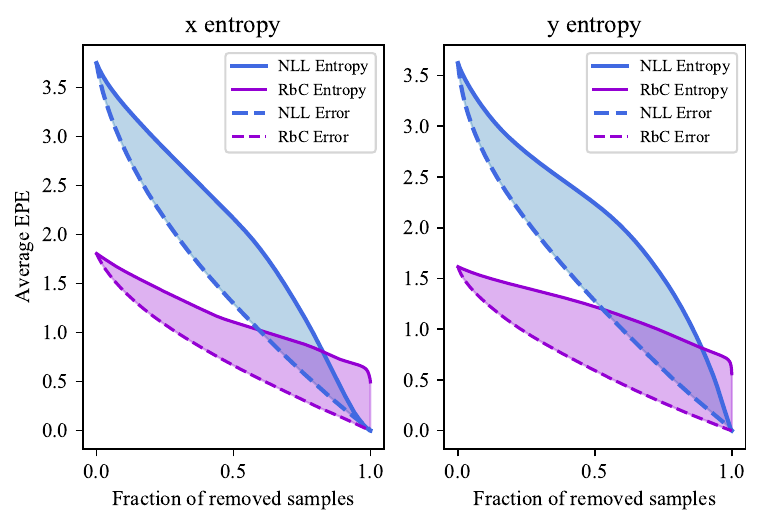}  
  \captionsetup{labelformat=empty}
  \addtocounter{figure}{-1}\vspace{-5pt}
  \caption{(b)} \vspace{-5pt}% Positioning error cumulative distribution function when adjacent samples are separated with a distance $\frac{3}{4}\lambda$.}
  \label{NLoSAcc}
\end{subfigure}
\begin{subfigure}
  \centering
  % include second image
  \includegraphics[width=0.9\linewidth]{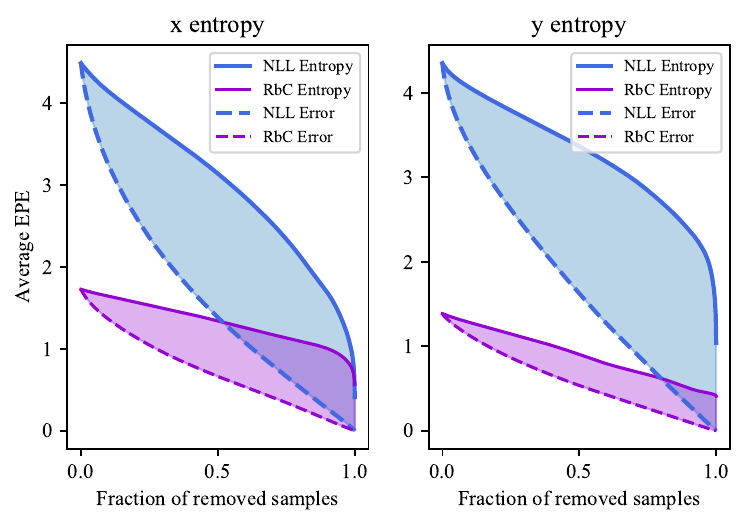}  
  \captionsetup{labelformat=empty}
  \addtocounter{figure}{-1}\vspace{-5pt}
  \caption{(c)} \vspace{-5pt}% Positioning error cumulative distribution function when adjacent samples are separated with a distance $\frac{3}{4}\lambda$.}
  \label{MixAcc}
\end{subfigure}
\caption{Sparsification curves of NLL and RbC methods under high training density (4 laps as training data) and across three scenarios: (a) LoS, (b) NLoS, (c) Mixed.}
\label{Spar} 
\end{figure}
Our approach starts with assessing the validity of the input channel snapshot, as outlined in Section.~\ref{Section3}.~B. The first criterion, related to the CTF matrix $\boldsymbol{\Xi}$, employs a cut threshold set at $3500$ out of $5888$ ($128 \times 46$) available physical resource elements, approximately $60\%$, so that the channel information is sufficient. After discarding snapshots with insufficient data, we generate the amplitude of impulse response beam matrix  $|\mathbf{G}_t|$ and feed it to the attention-aided localization block. This block, with detailed parameters in Table.~\ref{table: NLL1}, comprises three cascaded sub-blocks. Initially, positioning encoding is applied to $|\mathbf{G}_t|$ using (\ref{PosE}). Subsequently, a layer normalization procedure follows according to (\ref{L7}). The normalized matrix is then input into a simple $2$-head self-attention block with a single self-attention layer, generating matrix $\mathbf{Z}'$ via (\ref{eq2}-\ref{Mul}). After the Add \& Norm operation, the output is transferred to the second sub-block, consisting of two position-wise Fully Connected Neural Networks (FCNNs) with sizes $\mathbf{W}_1 \in 46 \times 128$ and $\mathbf{W}_2 \in 128 \times 46$. Following this, the output matrix of the second sub-block is vectorized to yield a vector of length $5888$. This vector is fed into the last FCNN sub-block, with sizes as given in Table~\ref{table: Block3}. We compare the localization performance when using three different loss functions and in three typical scenarios.  %Our ML-based localization approach trains on only valid channel CSI, i.e., the Delay-Beam diagram.  In alignment with our data-cleaning framework, our localization approaches consist of three intermediate blocks as well. %Notably, the structures of the data-cleaning and localization networks differ solely in the third blocks. 
%Specifically, their structures and parameter settings are illustrated in Table~\ref{table: Block3}. 
As illustrated, the output matrix of the second intermediate block is first vectorized and fed to the input layer of the third sub-block, which consists of $2$-$3$ FCNNs depending on the choice of loss functions. When the loss function RbC is used, its corresponding network delivers the probability of all $L_x$ and $L_y$ bins. In scenario I, $L_x = L_y = 200$ while in the other two scenarios $L_x = L_y = 100$. The deviation vectors $\mathbf{d}_x$ and $\mathbf{d}_y$ are set as: $\mathbf{d}_x = \delta_x \mathbf{1}, \mathbf{d}_y = \delta_y \mathbf{1}$, where $\mathbf{1}$ denotes the all-ones vector, $\delta_x$ and $\delta_y$ denote the deviation value of the $x$- and $y$-axis, respectively. Accordingly, the output dimension $\tilde{L}$ equals $L_x +L_y +2$. The penalty term $\gamma_1$ and $\gamma_2$ are set as: $\gamma_1 = \gamma_2 = 1$. In addition, the Euclidean norm loss function is utilized, i.e. $\eta = 2$. 

Fig. \ref{Single_SP_PoS} compares the positioning accuracy of our single-snapshot localization pipeline using three loss functions in three scenarios under different training densities. As shown, the RbC method outperforms the other two methods in all three scenarios and under both high and low training densities. Compared to the other two methods,  RbC learns better the non-Gaussian probability distribution of the UE position, while the performance of the NLL method is constrained by its underlying Gaussian assumption, and the MSE method does not estimate uncertainty. The performance of these three methods differ less in the LoS scenario and high training density, because the estimated UE position has less uncertainty in this situation. However, in other scenarios or lower training density, the uncertainty of the estimated UE position increases due to reduced SNR or training samples. Consequently, an accurate uncertainty estimation is more essential, and thus the RbC method performs much better. At both high and low training densities, our pipeline performs best in LoS scenarios, the mixed scenario ranks  $2$nd, while the localization performance in the NLoS scenario is the worst. We postulate that in the LoS scenario, the much higher SNR contributes to very good positioning accuracy. 

To further compare the uncertainty estimation quality of the NLL and RbC methods, we demonstrate the sparsification and oracle curves of the probability density functions of the estimated UE-$x$ and $y$ coordinates under high training density in Fig.~\ref{Spar}. Specifically for the NLL method, we discretize the predicted Gaussian functions to achieve the same number of discrete bins as the RbC method, according to \eqref{DGaussain}. The AUSE values for all training densities are calculated according to \eqref{EqAUSE} and are displayed in Table~\ref{table: AUSE}. To reduce the effect of outliers, the starting point of the sparsification and oracle curves equals $99\%$ of the positioning error. As depicted in Fig.~\ref{Spar}, the discrepancies between the sparsification (entropy) and oracle curves are significantly reduced in all three scenarios when the RbC method is used. This improvement is reflected in the improved AUSE values presented in Table~\ref{table: AUSE}. These findings underscore the quality of the uncertainty estimation achieved with our approach.

\begin{table}
\centering
\caption{Structures and parameter settings of the third FCNN sub-block using three different loss functions.}%, the numbers in parentheses represent the distances between two training samples
{\begin{center}
\begin{tabular}{ccccccc}
\hline\hline
\backslashbox[30mm]{\small Items}{{\small Loss F.}} & MSE & NLL & RbC\\
\hline
Input layer size & $5888 \times 1$ & $5888 \times 1$ & $5888 \times 1$\\
Hidden layer 1 & $5888 \times 128$ & $5888 \times 128$ & $5888 \times 128$ \\
Hidden layer 2 & $128 \times 32$ & $128 \times 32$& \hspace{-7pt}$128\times \tilde{L}$\\
Hidden layer 3 & $32 \times 2$ & $32 \times 4$ & N/A  \\
Batch size & $64$ & $64$& $64$  \\ 
Lr: LoS (4 laps) & $0.0006$ & $0.0006$ & $0.0006$  \\
Lr: NLoS (4 laps) & $0.0006$ & $0.0006$ & $0.0006$  \\
Lr: Mixed (4 laps) & $0.0006$ & $0.0006$ & $0.0006$  \\
Lr: LoS (2 laps) & $0.0002$ & $0.0002$ & $0.0002$ \\
Lr: NLoS (2 laps) & $0.0001$ & $0.0001$ & $0.0001$  \\
Lr: Mixed (2 laps) & $0.0002$ & $0.0002$ & $0.0002$  \\
Learning Epoch  & $100$ & $100$ & $100$\\
Cost function & \eqref{MSE} &\eqref{NLL} & \eqref{RbC}\\
\hline\hline
\end{tabular}
\end{center}
}
\label{table: Block3}
\end{table}

\begin{table}[t!]
\centering
\caption{AUSE values of two uncertainty estimation algorithms under different training densities across three channel scenarios.}%, the numbers in parentheses represent the distances between two training samples
{\begin{center}
\begin{tabular}{ccccccc}
\hline\hline
 & NLL-x & RbC-x & NLL-y & RbC-y\\
\hline
LoS (4 laps) & 0.480 & 0.179 & 0.351 & 0.163 \\
NLoS (4 laps) & 0.579 & 0.427 & 0.704 & 0.548 \\ 
Mixed (4 laps) & 1.428 & 0.616 & 1.543 & 0.325\\
LoS (2 laps) & 1.951 & 0.968 & 2.023 & 1.181  \\ 
NLoS (2 laps) & 3.816 & 1.868 & 3.407 & 2.475  \\
Mixed (2 laps) & 4.682 & 0.809 & 3.540 & 1.138\\
\hline\hline
\end{tabular}
\end{center}}
\label{table: AUSE}
\end{table}

\subsection{Smoothing the trajectory by Kalman filtering}
\begin{figure}[h]
  \centering
\begin{subfigure}{}
  \centering
  % include first image
  \includegraphics[width=1.0\linewidth]{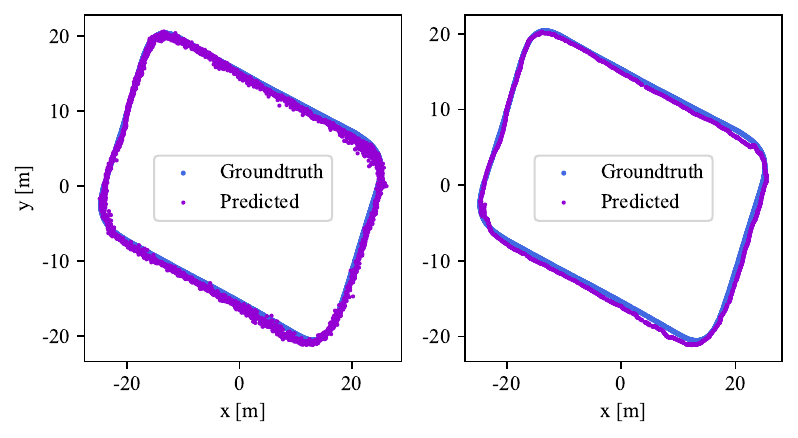}  
  \captionsetup{labelformat=empty}
  \addtocounter{figure}{-1} \vspace{-10pt}
  \caption{(a)} \vspace{-5pt}%Positioning error cumulative distribution function when adjacent samples are separated with a distance $\frac{1}{8}\lambda$.}
  \label{LoSAcc}
\end{subfigure}
\begin{subfigure}
  \centering
  % include second image
  \includegraphics[width=1.0\linewidth]{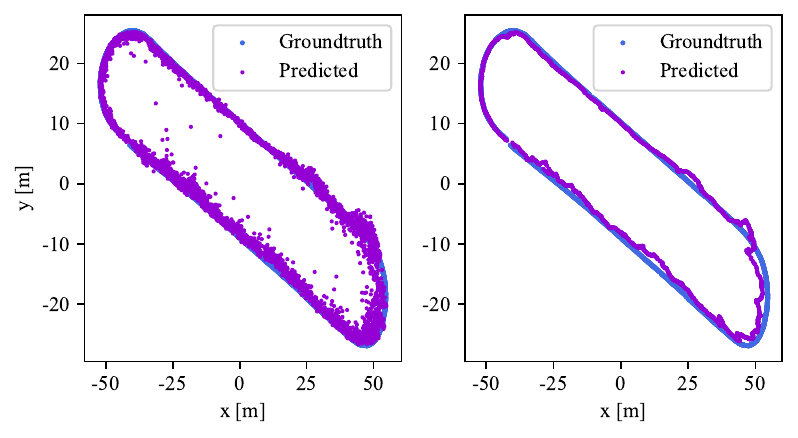}  
  \captionsetup{labelformat=empty}
  \addtocounter{figure}{-1}\vspace{-10pt}
  \caption{(b)} \vspace{-5pt}% Positioning error cumulative distribution function when adjacent samples are separated with a distance $\frac{3}{4}\lambda$.}
  \label{NLoSAcc}
\end{subfigure}
\begin{subfigure}
  \centering
  % include second image
  \includegraphics[width=1.0\linewidth]{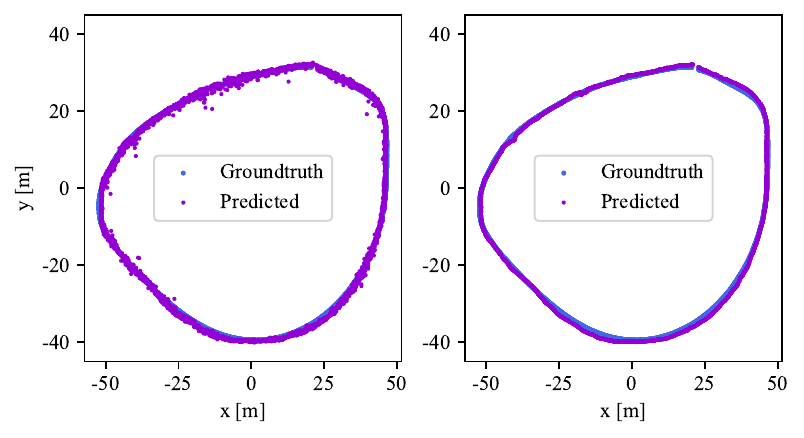}  
  \captionsetup{labelformat=empty}
  \addtocounter{figure}{-1}\vspace{-10pt}
  \caption{(c)} \vspace{-5pt}% Positioning error cumulative distribution function when adjacent samples are separated with a distance $\frac{3}{4}\lambda$.}
  \label{MixAcc}
\end{subfigure}
\caption{Comparison between the raw (the left) and Kalman-filtered trajectory (a) LoS, (b) NLoS, (c) Mixed.}
\label{Traj} \vspace{-3pt} 
\end{figure}
Next, we investigate the performance when using a Kalman filter for smoothing within our pipeline. To clearly visualize the effect of the Kalman filter, we apply a low training density, using two laps for training and one lap for testing. First, the validity of each channel CSI is assessed by the data cleaning block. All test channel samples classified as valid are then utilized for evaluation. Similarly to Section \ref{Section5}.B, we apply an attention-aided block as the backbone and the output layer utilizes the RbC uncertainty estimation. For simplicity, the matrix $\boldsymbol{\Lambda}$ in \eqref{XiGamma} and the matrix $\mathbf{R}$ in \eqref{GammaR} are set as 
\begin{equation}
    \boldsymbol{\Lambda} = \epsilon_{1}^2\hspace{1pt}\mathbf{I}, \hspace{5pt} \mathbf{R} =  \epsilon_{2}^2\hspace{1pt}\mathbf{I}, \vspace{-2pt}
\end{equation}
where $\epsilon_{1}$ and $\epsilon_2$ denote the standard deviation, which indicates the state and observation noise levels, respectively. Their exact values for the three scenarios are listed in Table~\ref{table: KF}. 
\begin{table}[t!]
\centering
\caption{Parameter settings and performance evaluation when applying the Kalman Filtering}%, the numbers in parentheses represent the distances between two training samples
{\begin{center}
\begin{tabular}{ccccccc}
\hline\hline
 & $\epsilon_1$ & $\epsilon_2$ & RMSE (m), before filter & RMSE (m)\\
\hline
LoS  & 0.05 & 1.2 & 0.99 & 0.92  \\ 
NLoS & 0.05 & 1.2 & 2.00 & 1.75  \\
Mixed & 0.05  & 1.2 & 1.01 & 0.82 \\
\hline\hline
\end{tabular}
\end{center}}
\label{table: KF}
\vspace{-10pt}
\end{table}
Fig.~\ref{Spar} shows the predicted UE trajectories both with (right) and without (left) the Kalman filter for the three scenarios. The MSE between the predicted trajectories and their ground truths is shown in Table.~\ref{table: KF}. As expected, the results demonstrate a significant improvement with the inclusion of the Kalman filter: the trajectories become considerably smoother, and outliers are mitigated to a large extent. Consequently, there is a substantial enhancement in localization accuracy, particularly evident in NLoS and mixed propagation scenarios. This improvement can be attributed to the ability of the Kalman filter to utilize relationships between different snapshots, which effectively balances the newly predicted UE position with previous positional states, leading to more accurate localization.

\section{Conclusions}
%This paper demonstrates how an attention-enabled generative AI model can be utilized for outdoor user positioning in a commercial TDD 5G NR system. The presented results outperform existing outdoor user localization studies and show sub-1 m accuracy. On the last note, further studies are recommended to increase the localization accuracy by using a higher sampling rate of the SRS channel estimates.
\label{Section6}
In this paper, machine learning is applied to a 5G NR cellular system for UE localization. A novel ML-based localization pipeline is presented, which utilizes attention-aided techniques to estimate UE positions by employing impulse response beam matrices as channel fingerprints. In addition, we implement two uncertainty estimation techniques, namely the NLL and RbC methods, to estimate the probability density function of the UE position error and compare their performances. Finally, a Kalman filter is applied to smooth consecutive position estimates. To evaluate our pipeline, an outdoor cellular 5G measurement campaign was conducted at 3.85 GHz with a 100 MHz bandwidth, covering both LoS and NLoS scenarios, achieving submeter-level localization accuracy. The measurement results indicate several key findings: 1) The attention-aided block shows promising potential to deliver high-precision localization accuracy. 2) The RbC uncertainty method outperforms the traditional NLL method, particularly with low training density or in more complex channel propagation scenarios. This advantage likely stems from the fact that the RbC method is not constrained by a Gaussian assumption on position errors. 3) Applying a Kalman filter to smooth consecutive position estimates significantly reduces position outliers, thereby enhancing localization accuracy.

\section*{ACKNOWLEDGMENT}
The authors thank PhD candidate Ziliang Xiong for valuable suggestions regarding the uncertainty prediction.
\bibliographystyle{IEEEtran}
\bibliography{reference}

\vfill\pagebreak

\end{document}